\documentclass[11pt]{article}

\usepackage{amsbsy}
\usepackage{amsmath}
\usepackage{multirow}
\usepackage{graphicx}

\begin{document}

\large\textit{\textbf{Combining Empirical Likelihood and Robust Estimation Methods for Linear Regression Models}}
\vspace{0.25in}

\textsuperscript{a}\c{S}enay \"{O}zdemir and \textsuperscript{b}Olcay Arslan
\vspace{0.25in}

\textsuperscript{a}Department of Statistics, Afyon Kocatepe University, 03200, Afyonkarahisar, Turkey, senayozdemir@aku.edu.tr

\textsuperscript{b}Department of Statistics, Ankara University, 06100, Ankara, Turkey
\vspace{0.25in}

\textit{\textbf{Abstract}}

 Ordinary least square (OLS),  maximum likelihood (ML) and  robust methods are the widely used methods to estimate the parameters of a linear regression model. It is well known that these methods perform well under some  distributional assumptions on  error terms. However,  these distributional  assumptions on the errors may not be appropriate for some data sets. In these case, nonparametric  methods  may be considered to carry on the regression analysis. Empirical likelihood (EL) method is one of these nonparametric methods.  The EL method maximizes a function, which is multiplication of the unknown probabilities corresponding to each observation, under some constraints  inherited  from  the normal equations in OLS estimation method.  However, it is well known that the OLS method has poor performance when there are some outliers in the data.  In this paper, we consider the EL method with robustifyed constraints. The robustification of the constraints  is done by using the  robust M estimation  methods  for regression.  We provide a small simulation study and a real data example to demonstrate the capability of the robust EL method to handle unusual observations in the data. The simulation and real data results reveal that robust constraints are needed when heavy tailedness and/or outliers are possible in the data.

\vspace{0.25in}

\textit{\textbf{Keywords:}}Empirical likelihood; Linear regression; M estimation
\vspace{0.25in}

\textit{\textbf{Class codes:}}
\vspace{0.25in}


\section{Introduction}

Consider the linear regression model

\begin{equation}\label{1}
    Y_{i}=\boldsymbol{{X_{i}^{T}}{\beta}}+\varepsilon_{i} \quad \textrm{for}\quad i=1,2,\ldots,n
\end{equation}

\noindent where $Y_{i}$  $\in R$ is the response variable, $\mathbf X_{i}\in R^{k}$ is the $k-$dimensional vector of the explanatory variables, $\boldsymbol\beta \in R^k$ is an unknown $k-$dimensional parameter vector and $\varepsilon_{i}$'s are the independent and identically distributed (iid) errors with $E(\varepsilon_{i})=0$ and $Var(\varepsilon_{i})=\sigma^2$. The regression equation given in (\ref{1}) can also be written in matrix notation as

\begin{equation*}
Y=\boldsymbol{X}{\boldsymbol{\beta }}+{\boldsymbol{\varepsilon}},
\end{equation*}

\noindent where $\boldsymbol{X}_{n\times k}$ is the design matrix, $Y$ is the response vector, and $\boldsymbol{\varepsilon }$ is the vector of $\varepsilon _{i}.$

The  simplest way to estimate the parameters of a linear regression model is to use the OLS method. The OLS estimators for $\boldsymbol{\beta}$ can be obtain by minimizing the following objective function:

\begin{equation}\label{2}
\frac{1}{n}\sum_{i=1}^{n}\left(Y_{i}-\boldsymbol{ X_{i}^{T}\beta}\right)^{2}.
\end{equation}

\noindent Taking the derivative of this function with respect to $\boldsymbol{\beta}$  and setting to zero gives the following estimating equation, which is known as the normal equations in the OLS:

\begin{equation}\label{3}
\frac{1}{n}\sum_{i=1}^{n}\boldsymbol X_{i}(Y_{i}-\boldsymbol {X_{i}^{T}\beta})=0.
\end{equation}

\noindent From this equation  the OLS  estimator $\hat{\boldsymbol\beta}$ is obtained. An unbiased estimator for  $\sigma^2$  can be obtained  using   residuals sum of squares  after the OLS estimator for the regression parameter  is obtained.  The following  objective function, which is the negative of the log-likelihood function  under normally distributed error terms,can also be minimized to obtain estimators for   $\boldsymbol\beta$  and $\sigma^2$.

\begin{equation}\label{4}
\frac{1}{2\sigma^2}\sum_{i=1}^{n}(Y_{i}-\boldsymbol {X_{i}^{T}\beta})^{2} + nlog\sigma.
\end{equation}

\noindent The ML  estimators  will be the solutions of the equation given in (\ref{3}) and  the following equation

\begin{equation}\label{5}
\frac{1}{n}\sum_{i=1}^{n}((Y_{i}-\boldsymbol X_{i}^{T}\beta)^{2}-\sigma^2)=0.
\end{equation}

\noindent Note that the OLS  and the ML estimators for the regression parameters are the same under the normality assumption on error.

It is known that the OLS estimators (also the ML estimators) are the minimum variance unbiased estimates for model parameters under the assumption that $\varepsilon_{i}$ are normally distributed. However, these estimators are dramatically  effected  when fundamental assumptions are unfulfilled by the nature of the data. It is well known that OLS estimators (or ML estimators under the normality assumption)  are very sensitive to outliers or departure from normality ( such as heavy-tailed error).  Even  a single  outlier may drastically  affect  the OLS estimators.  Therefore,  robust regression estimation methods  alternatives to the OLS estimation method  have been developed  to deal with outliers and/or heavy-tailed error distributions.  In this paper,  we will use the M regression estimation method  (Huber (1964)) that will be briefly described  in the following paragraph. One can see the books by Huber(1981) for the details of the robust regression estimation  methods,  including M estimation methods.

The widely  used method of robust regression is the M-estimation   introduced by Huber (1964,1973). Since, this class of estimators can be regarded as a generalization of maximum-likelihood estimation the  term M estimation is used.    The   M estimation method is designed to minimize an   objective function that is less  rapidly increasing  then the OLS objective  function.  This is done using a $\rho$  function  that is less rapidly incising than the squared function.   The M estimator for the regression parameter $\beta$ is obtained by minimizing the following objective function with respect to $\beta$.

\begin{equation}\label{6}
\frac{1}{n}\sum_{i=1}^{n}\rho(Y_{i}-\boldsymbol{X_{i}^{T}\beta}).
\end{equation}

\indent The function $\rho$   should be nonnegative,  nondecreasing and $\rho(0)=0$. The M estimation method   includes  OLS,  ML and  Least absolute deviation (LAD) methods  with the choice of   $\rho(t)=t^{2},   -\log(f(t))$, and $|t|$, respectively.  In this minimization problem,  contribution of each observation to  the sum  is determined by the function $\rho$ in the sense that  the  observations with large residuals have small contributions to the sum.   Consequently, since  the outlying observation will   have small  contribution to the sum due to the $\rho$ function  they will  not drastically effect the resulting   estimator.  If $\rho$ is differentiable, differentiating (\ref{6})   with respect to $\beta$ and setting to zero yields the following  M estimation equation

\begin{equation}\label{7}
\frac{1}{n}\sum_{i=1}^{n}\psi(Y_{i}-\boldsymbol{X_{i}^{T}\beta})\boldsymbol{X_{i}}=0,
\end{equation}

\noindent   where $\psi=\rho'$.

Note that in  robust statistical analysis, the influence function  is one way   to measure the robustness of an estimator and it is desired to have a bounded influence function.  In M estimation method the function $\psi$ determines the shape of the influence function. Therefore, if $\psi$  is nonincreasing  (may be equal to zero beyond some threshold  or tend to zero in larger values of its argument) the corresponding M estimator will  have bounded influence  function.  In regression M estimation bounded $\psi$  function only  controls the large residuals; that is,  the estimator will be resistent to the outliers in  y direction.  To   control the outliers in x-direction we have to use other robust   estimation   methods such as   generalized M  estimation or MM regression estimation methods.   In this paper,  we will  only combine  the regression M estimation with the EL method and save the   others for our next project.  In robust statistics literature   several  different $\rho$  functions are proposed,  but  Huber  and Tukey (bisquare) $\rho$  functions are the widely  used ones.  The Huber $\rho$  and   $\psi$  functions are

\begin{equation*}
\rho=\begin{cases}
           \left(Y_{i}-\boldsymbol{X_{i}^{T}\beta}\right)^{2} & ,|Y_{i}-\boldsymbol{X_{i}^{T}\beta}|\leq k \\
           2k|Y_{i}-\boldsymbol{X_{i}^{T}\beta}|-k^{2} & ,|Y_{i}-\boldsymbol{X_{i}^{T}\beta}|>k
            \\
         \end{cases}
\end{equation*}

\begin{equation*}
\psi=\begin{cases} \left(Y_{i}-\boldsymbol{X_{i}^{T}\beta}\right) & ,|Y_{i}-\boldsymbol{X_{i}^{T}\beta}|\leq k \\
           sgn\left(Y_{i}-\boldsymbol{X_{i}^{T}\beta}\right)k & ,|Y_{i}-\boldsymbol{X_{i}^{T}\beta}|>k
            \\
\end{cases}
\end{equation*}

\noindent Similarly,  The bisquare (Tukey) $\rho$  and   $\psi$  functions are

\begin{equation*}
\rho=\begin{cases}
           \frac{k^{2}}{6}\left(1-\left(1-\left(\frac{Y_{i}-\boldsymbol{X_{i}^{T}\beta}}{k}\right)^{2}\right)^{3}\right) & ,|Y_{i}-\boldsymbol{X_{i}^{T}\beta}|\leq k \\
           \frac{k^{2}}{6} & ,|Y_{i}-\boldsymbol{X_{i}^{T}\beta}|>k
            \\
         \end{cases}
\end{equation*}

\begin{equation*}
\psi=\begin{cases}
\left(1-\left(\frac{Y_{i}-\boldsymbol{X_{i}^{T}\beta}}{k}\right)\right)^{2}\left(Y_{i}-\boldsymbol{X_{i}^{T}\beta}\right) & ,|Y_{i}-\boldsymbol{X_{i}^{T}\beta}|\leq k\\
      0 & ,|Y_{i}-\boldsymbol{X_{i}^{T}}\beta|>k
\end{cases}
\end{equation*}

Concerning the estimation of   $\sigma^{2}$    along  with  $\boldsymbol\beta$  the  following objective   function can be minimized to obtain  M estimators for  $\boldsymbol\beta$  and $\sigma^2$  (see Huber and Ronchetti, 2009, Chapter 7.7).

\begin{equation}\label{11}
\frac{1}{n}\sum_{i=1}^{n}\rho\left(\frac{Y_{i}-\boldsymbol{X_{i}^{T}\beta}}{\sigma}\right)\sigma+a\sigma,
\end{equation}

\noindent  Taking the derivative of this objective function  and setting to  zero  yield  the following M estimating   equations   for  $\boldsymbol\beta$  and $\sigma^2$

\begin{equation}\label{12}
\frac{1}{n}\sum_{i=1}^{n}\psi\left(\frac{Y_{i}-\boldsymbol{X_{i}^{T}\beta}}{\sigma}\right)\boldsymbol X_{i}=0,
\end{equation}

\begin{equation}\label{13}
\frac{1}{n}\sum_{i=1}^{n}\rho_{0}\left(\frac{Y_{i}-\boldsymbol{X_{i}^{T}\beta}}{\sigma}\right)=a
\end{equation}

\noindent where $\rho_{0}(r) =r\psi(r)-\rho(r)$.   $a>0$ is added in order to get consistency  of the scale estimator under normality assumption of the error and to get the classical  OLS estimator for the scale when $\rho(x)=x^{2}/2$. Note that, since throughout this study  we are only interested  for estimating  the regression parameters  we can assume that $\sigma$ is fixed.  This can be done either by estimating $\sigma$ beforehand or assuming that it is known, as it is suggested in the book by Maronna et al. (2006) (Chepters 4.4.1  and4.4.2.).

One can see the books Huber (2009) and Maronna (2006) for further  details on  M estimators for regression and its properties.

If  we do not want to use OLS or robust methods  or not want to assume any distribution for the error terms we can also use nonparametric methods as alternatives to  these mentioned methods  for estimation and inference in a linear regression model. One of these nonparametric methods is the EL method  which was introduced by Owen(1988). This method is an alternative to likelihood method when there is no distribution assumption for error terms. In EL method, it is assumed that each observation has unknown $p_{i}$ probability weight for $i=1,2,\ldots,n$. The aim of the EL method is to estimate the regression parameters by maximizing an empirical likelihood function defined as the multiplication of these $p_{i}$ s' , which is developed modifying  nonparametric likelihood ratio function, under some constraints  related to the parameters of interest. However, since the constraints used in EL  method (see Section 2)  are   very similar  to   the normal equations in   OLS or  likelihood equations under normally distributed errors  the corresponding estimators will be very sensitive to the non-normality or unusual  observations (outliers)  in data.  In this paper,  we will carry on EL estimation in regression using robust constraint  borrowed  from the robust estimation equations   described in previous paragraph. Since,  we will   combine the M estimation methods with the EL method we expect that the   resulting   estimator for the regression parameters will be resistent  to the outliers   in y-direction.  Simulation study  and real  data example results  show  that this expectation comes true.

Note that the reason we keep $1/n$ multiplication in   all the objective  functions  and the estimating equations is to emphasize that each observation equally contributes ($1/n$) to the minimization procedure.   In the EL method, which will be  described  in Section 2, this term will be replaced by an unknown probability $p_{i}$ corresponding to each observation. Therefore, in empirical likelihood procedure   contribution of each observation will  be different  and this contributions (probabilities) need to be estimated.

The rest of the paper is organized as follows. In next section, after we briefly describe the EL method we will move on the EL estimation  with robust constraint for linear regression models. In Section 3 we will provide a small  simulation study and  a real data example to  illustrate the necessity  of robust constraints in EL  estimation. The paper is finalized  with a conclusion section.

\section{Empirical Likelihood  Estimation for the Parameters of a Linear Regression Model with Robust Constraints}

In this section,   we will first outline   the  EL method with classical constraints.    Then, we will move on our proposal that combines   the EL estimation method  with some robust constraints.

\subsection{Empirical Likelihood  Estimation}
Consider the linear regression model given in equation (\ref{1})  with the same assumptions  given there. The EL estimators will be the solution of the   the following constrained optimization problem.  Specifically, if we are only interested for estimating the regression parameter $\boldsymbol\beta$ the EL method maximizes the  following empirical likelihood function

\begin{equation}\label{14}
L(\boldsymbol\beta)=\prod_{i}^{n}p_{i}
\end{equation}

with  respect to $p_{i}\geq 0$ and $\boldsymbol\beta$  under the  following constraints

\begin{equation}\label{16}
\sum_{i}^{n}p_{i}=1
\end{equation}

\begin{equation}\label{17}
\sum_{i}^{n}p_{i}\left(Y_{i}-\boldsymbol{X_{i}^{T}\beta}\right)X_{i}=0
\end{equation}

\noindent where $ p_{1},p_{2},\dots,p_{n}$ are the probability weights of the observations.  Note that the constraint given in equation (\ref{17}) is very similar to the normal equation given in (\ref{3}).  The only difference is  the unknown probability weights assigned for each observation. In former equation each observation has equal probability $1/n$, but in later each observation has a different $p_{i}$ probability  and these probabilities are unknown and  need to be estimated. Further,  if  we are interested  for estimating $\sigma^2$ along with the regression parameters the following constraint, which is  motivated  form the equation (\ref{5}), is  added to   constrained optimization problem defined above.

\begin{equation}\label{18}
\sum_{i}^{n}p_{i}[(Y_{i}-\boldsymbol{X_{i}^{T}\beta})^{2}-\sigma^{2}]=0.
\end{equation}

The empirical likelihood method can be used estimating parameters, constructing confidence regions, testing statistical hypothesis, etc. Briefly, it is a useful tool for making statistical inference when it is not too easy to assign a distribution to data. There are several remarkable studies on the EL method after it was introduced by Owen (1988, 1990, 1991). In these papers, Owen   used the empirical likelihood for constructing confidence regions and estimating parameters of linear regression models. Hall and Scala (1990) studied on main features of the empirical likelihood, Kolaczyk (1994) adapted it in generalized linear regression model, Qin and Lawless (1994) combined general estimating equations and the empirical likelihood, Chen et. all (1993,1994 1996, 2003,2009) handled this method for constructing confidence regions, parameter estimation with additional constraints, Newey and Smith  (2004) studied about properties of generalized methods of moments and generalized empirical likelihood estimators, Bondell and Stefanski (2013) suggested a robust estimator modified the generalized empirical likelihood. Recently, Ozdemir and Arslan (2017) have suggested an  alternative   algorithm to obtain EL  estimates  using the primal optimization problem. As we have already  pointed out we are proposing to assign robust constraints   to gain  robustness.

The estimation of $p_{i}$s and the model parameters $\boldsymbol\beta$ and $\sigma^{2}$ can be done by maximizing the function given in equation (\ref{14}) under the constraints \ref{16}  and \ref{17}. For simplicity we assume that variance is known  and consider only estimation of $\boldsymbol\beta$.   Therefore, the EL  function should be jointly maximized with respect to $p_{i}$s  and $\boldsymbol\beta$. One way to handle this constraint maximization problem is  as follows.  First, fix the regression parameter  vector $\boldsymbol\beta$ and consider maximizing  the log-EL  function  with respect to $[p_{1},p_{2},\ldots,p_{n}]^{T}$  under the constraints given above. This procedure is called as  profiled out. Once this is done, the profile likelihood  will be a function of $\boldsymbol\beta$.  Then,  we can maximize the profile likelihood to obtain  the EL  estimator of $\boldsymbol\beta$. This problem will be easily   handled  using Lagrange multiplier  method.   Therefore, setting the problem in Lagrangian form, the Lagrange function associated  with this constrained maximization  problem is

\begin{equation}\label{19}
    L(\boldsymbol{p,\beta},\lambda_{0},\boldsymbol\lambda_{1}^{T})=\sum_{i=1}^{n}log(p_{i})-\lambda_{0}(\sum_{i=1}^{n}p_{i} - 1)-n\boldsymbol\lambda_{1}^{T}\left(\sum_{i=1}^{n} p_{i}\boldsymbol{X_{i}}\left(Y_{i}-\boldsymbol{X_{i}^{T}\beta}\right)\right)
\end{equation}

\noindent where $\mathbf {p}=[p_{1},p_{2},\ldots,p_{n}]^{T}$   are the vector of probabilities, and  $\lambda_{0} \in R^{1}$ and $\boldsymbol\lambda_{1}^{T}\in R^{p}$ are the Lagrange multipliers.  Taking the derivatives of equation (\ref{19}) with respect to each $p_{i}$, and setting them to zero yields

\begin{equation}\label{20}
    p_{i}=\frac{1}{n\left(1+\boldsymbol\lambda^{T}\boldsymbol{X_{i}}\left(Y_{i}-\boldsymbol{X_{i}^{T}\beta}\right)\right)},
\end{equation}

\noindent and $\boldsymbol\lambda=-n\boldsymbol\lambda_{1}$. By using equation (\ref{20}) in the log-EL function we find

\begin{equation}\label{21}
   L(\boldsymbol{\beta,\lambda})= -\sum_{i=1}^{n}\log\left(1+\boldsymbol{\lambda^{T}X_{i}}\left(Y_{i}-\boldsymbol{X_{i}^{T}\beta}\right)\right)-n\log n.
\end{equation}

For  a given $\boldsymbol{\beta}$  the Lagrange multiplier $\boldsymbol\lambda_{1}(\boldsymbol{\beta})$  will  be obtained  as the solution of the following  minimization problem

\begin{equation}\label{22}
  \boldsymbol\lambda_{1}(\boldsymbol{\beta})= argmin_{\boldsymbol{\lambda}_{1}}\left(-L(\boldsymbol{\beta},\boldsymbol{\lambda}_{1})\right).
\end{equation}

\noindent  Note that since this minimization problem will not have an explicit solution  numerical  methods  should be used to find the minimizer.  Now, using this in equation \label{21}   yields  the  following   function

\begin{equation}\label{23}
   L(\boldsymbol{\beta})= -\sum_{i=1}^{n}\log\left(1+\boldsymbol\lambda_{1}(\boldsymbol{\beta})^{T}X_{i}\left(Y_{i}-\boldsymbol{X_{i}^{T}\beta}\right)\right)-n\log n.
\end{equation}

\noindent Then, this function  will  be   maximized  to   obtain  the EL  estimator   for the regression coefficient  vector $\boldsymbol{\beta}$.    That is the EL estimator of $\boldsymbol{\beta}$  will  be

\begin{equation}\label{24}
  \hat{\boldsymbol{\beta}}_{EL}= argmax_{\boldsymbol{\beta}}L(\boldsymbol{\beta}).
\end{equation}

\noindent Again, numerical algorithms will be required to handle this maximization problem.    Note that  one can see   Owen(1989,2001)or Kitamura (2006)  for further   details about the computational issues of the EL  method.   In the following subsection we will  use  similar steps  to   maximize the EL  function under robust  constraints.

\subsection{Empirical Likelihood Estimation  with Robust Constraints}

In this subsection   we will turn our attention  to the EL estimation using   robust constraints instead of the classical constraints.  In the EL method  the   unknown probabilistic weights $p_{i}$  are assigned for each observation  so that each observation will  have different  contribution to the estimation procedure. In some sense,  this  can be considered  as a weighting procedure of the normal equations in terms of probabilities for each observation. However, from our limited experience we observe that the probabilistic  weights are not satisfactorily   reduce the  effect of outliers on the estimators. Therefore,  extra care should be taken to reduce  the outliers  affect on estimation procedure. This   can be done by adapting the M estimating equation given in previous section.

Suppose that we are  interested   for  estimating  the regression parameter $\boldsymbol{\beta}$. Then,   we will  maximize  the following EL function

\begin{equation}\label{23}
L_{EL}(\beta)=\prod_{i=1}^{n}p_{i}
\end{equation}

\noindent under the  constraints

\begin{equation}\label{24}
\sum_{i=1}^{n}p_{i} = 1
\end{equation}

\noindent and

\begin{equation}\label{25}
\sum_{i=1}^{n}p_{i}\psi\left(Y_{i}-X_{i}^{T}\beta\right)X_{i}=0,
\end{equation}

\noindent where  $p_{i}\geq 0.$ The second   constraint  can be regarded  as robust version  of the classical   constraint  related    to  the regression parameters. Here,   $\psi$ function is a  nonincreasing function of the residuals for Huber case  and a decreasing   function of   residuals   for Tukey   function.  Thus,   unusual observations with large   residuals  will  receive   small $\psi$ values   so that the corresponding observations will  not completely ruin   the  estimation procedure.  Further, if we are also interested for  estimating  $\sigma^2$ along  with the  regression  parameters the simultaneous M estimating equations given in  equations (\ref{12})-(\ref{13})  can  be   adapted and used in the maximization   problem  instead of  the classical   constraints for  $\beta$ and $\sigma^2$  given in  equations (\ref{17})and (\ref{18}). The adaptation of those  equations    will be done  using  $p_{i}$'s  instead  of  using  $1/n$. However,   since in this paper  we are only interested  for the regression parameters we will  solve the above maximization   problem and   not  add  a robust  constraint for $\sigma^2$.

The robust EL  estimator for $\boldsymbol{\beta}$ will be defined  as the value of $\boldsymbol{\beta}$  that maximizes  the log-EL  function  under the  constraints   given in equations (\ref{24})  and (\ref{25}).  We will  again  use the  maximization procedure   described  in previous subsection. In this case, the  Lagrange function     will  be

\begin{equation}\label{26}
    L(\boldsymbol{p,\beta},\lambda_{0},\boldsymbol\lambda_{1}^{T})=\sum_{i=1}^{n}log(p_{i})-\lambda_{0}(\sum_{i=1}^{n}p_{i} - 1)-n\boldsymbol\lambda_{1}^{T}\left(\sum_{i=1}^{n}  p_{i}\psi\left(Y_{i}-\boldsymbol{X_{i}^{T}\beta}\right)X_{i}\right).
\end{equation}

\noindent Taking the derivatives of the Lagrange function  with  respect to  $p_{i}$, $\lambda_{0}$ and $\boldsymbol\lambda_{1}$, setting them to zero and solving the  corresponding first order  equations for this  maximization problem yield

\begin{equation}\label{27}
    p_{i}=\frac{1}{n\left(1+\boldsymbol\lambda^{T}\psi\left(Y_{i}-\boldsymbol{X_{i}^{T}\beta}\right)X_{i}\right)},  for \quad i=1,2,...,n,
\end{equation}

\noindent and $\boldsymbol\lambda=-n\boldsymbol\lambda_{1}$.   Then,  using these   $p_{i}$s    in the log-EL function  we get the  following objective function

\begin{equation}\label{28}
   L(\boldsymbol{\beta},\lambda)= -\sum_{i=1}^{n}\log\left(1+\boldsymbol\lambda^{T}\psi\left(Y_{i}-\boldsymbol{X_{i}^{T}\beta}\right)X_{i}\right)-n\log n,
\end{equation}

\noindent which is the function of $\boldsymbol\lambda$ and $\boldsymbol\beta$.    Further,   for a given regression parameter vector $\boldsymbol{\beta}$ the Lagrange multiplier $\boldsymbol\lambda$   can be obtained from  the following minimization problem

\begin{equation}\label{29}
  \boldsymbol\lambda(\boldsymbol{\beta})=argmin_{\boldsymbol\lambda}-\sum_{i=1}^{n}\log\left(1+\boldsymbol\lambda^{T}\psi\left(Y_{i}-\boldsymbol{X_{i}^{T}\beta}\right)X_{i}\right)-n\log n.
\end{equation}

\noindent Since, the  solution of this minimization problem  cannot be obtained explicitly,  numerical   algorithm  should be used to get the solution. Using $\boldsymbol\lambda(\boldsymbol{\beta})$,  the profile log-EL  function will  be

 \begin{equation}\label{30}
   L(\boldsymbol{\beta})= -\sum_{i=1}^{n}\log\left(1+\boldsymbol\lambda(\boldsymbol{\beta})^{T}\psi\left(Y_{i}-\boldsymbol{X_{i}^{T}\beta}\right)X_{i}\right)-n\log n.
\end{equation}

\noindent Then, the robust  EL  estimator   for the  regression parameter vector $\boldsymbol{\beta}$  will  be obtained as
\begin{equation}\label{22}
  \hat{\boldsymbol{\beta}}_{REL}= argmax_{\boldsymbol{\beta}}\left(L(\boldsymbol{\beta})\right).
\end{equation}
Again,  numerical algorithms  are necessary   to  preform  this  maximization  problem to  obtain  the robust  EL  estimates for the regression parameter   vector $\boldsymbol{\beta}$.

\section{Simulation Study}

To evaluate the performance of the proposed robust EL estimation procedure, we conduct a small simulation study.  We compare the classical  EL  method with the  robust   EL method (EL  with robust constraints) to  estimate  the parameters  of a  linear regression model.  We assume that  the error   variance is  known and  only deal with     estimating the regression parameters.  We take the  sample sizes as $n=30,50,100$  and  the dimensions  of the unknown   regression parameter  vector as $2,5,15$.  The  standard normal  distribution ($N(0,1)$) and the contaminated   normal  distribution ($(0.90)N(0,1)+(0.10)N(20,1)$)  are  used as the   error   distribution for the  regression model. The second error distribution is chosen  as to add  some  outliers in the data.  The  regressors   are also generated from the standard normal  distributions.   The unknown parameter vectors are also determined randomly from a normal distribution with mean $\mu$ and variance $\tau$.  The dependent variable is generated using the regression model given in equation (\ref{1}).  The MSE values and the relative efficiency of estimators with respect to the OLS are calculated to compare the performance of the   considered estimators.  All  the simulation  scenarios are repeated  100  times. The relative efficiencies are calculated using  the formula

\begin{equation*}
    RE=\frac{\sum_{j=1}^{100}\|\widehat{\boldsymbol\beta}_{j}-\boldsymbol\beta\|}{\sum_{j=1}^{100}\|\widehat{\boldsymbol\beta}_{j}^{OLS}-\boldsymbol\beta\|}
\end{equation*}

\noindent where $\widehat{\boldsymbol\beta}$ indicates the EL, EL-Hub, EL-Tukey, while $\widehat{\boldsymbol\beta}^{OLS}$ symbolizes the OLS.

The simulation results are summarized in Tables \ref{table1}-\ref{table4}. From these tables we observe that, without contamination all the estimators  have similar  performance with small MSE values  for the dimensions 2 and 5. When  $k=15$ the MSE values  of the estimators  are getting worse for  all the cases compare to  the  smaller dimension cases,  but  even for this case   the EL and  robust  EL  estimators   seem   superior   to the OLS estimators.  Smaller   behavior  is observed from the table of relative efficiencies.

In Tables \ref{table3} and \ref{table4}  we report  the simulation results for the contaminated   error distribution.   That is,    these   are the simulation   results for the  outlier  case.   From  these  result we observe that when we introduce contamination  the  robust EL  estimators  have better  performance  compare  to the  classical EL and the OLS estimators  in terms of the MSE  values.    Among   the two  robust   estimators   the Tukey    case  seems superior   to the Huber case in terms   of the MSE values.  Again,   when  the dimension  is large and  the  sample size is  small   all  the estimators  have large  MSE  values. However,  compare   to the classical  ones   the MSE values for the robust  estimators  are relatively  small. Overall,  from our limited simulation study  we observe that for all the simulation settings considered in this study, the robust  EL  estimators
have  comparable  performance in terms of MSE  values. Therefore, robust   constraints   should be considered   in  case of  potential     outliers   in the   data set.

\section{A real data example:International phone calls from Belgium }

\begin{figure}
  \includegraphics[width=\linewidth]{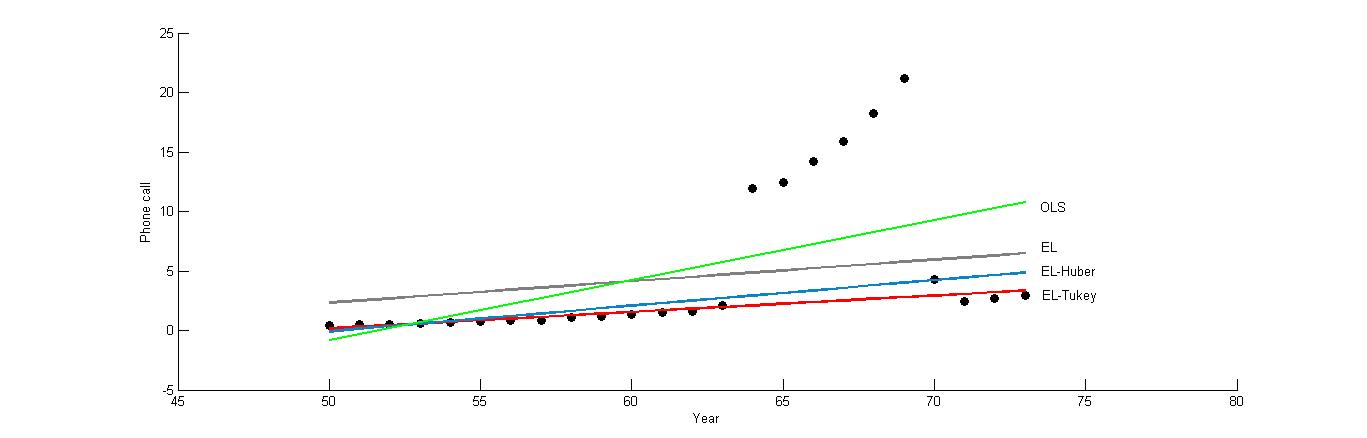}\\
  \caption{Number of international phone calls from Belgium in the years 1950-1973 with the OLS, EL, EL-Tukey, and EL-Huber fit}
  \label{Fig1}
\end{figure}

To illustrate proposed robust EL   estimators  we will  use a data set that is a widely used in  robust regression estimation  literature to  evaluate the performance  of the robust  estimation methods.  The data set, which is  taken from  the   book  by Rousseeuw and Leory (1987), contains of  the total number (in tens of millions) of international phone calls made over the years. Table \ref{table5}   displays the data set.  The scatter plot (Figure \ref{Fig1}) of the  data set shows an increasing  trend  over the years.  However,  it can be noticed  that  the data set  is heavily   contaminated  and these points  are the outliers in y-direction. Therefore,  using  classical  methods such as  OLS,   will  not provide adequate estimation. For instance,   the fitted line  obtained from the OLS method    is  highly  affected  from the contaminated   points (Figure \ref{Fig1}).   Therefore,  robust  regression estimation methods have been  used   to  get   better  fit  to this data set.  One can see  Rousseeuw and Leory (1987) for further details   and the results  of the robust  regression estimation methods.

Now  we will turn our attention   to the EL  estimation for the regression parameters.  If we   apply   classical   and robust   EL  estimation methods considered  in this paper    we   obtain the   regression estimates   provided   in Table  \ref{table6}.   This table also  contains the  OLS  estimates.   From   this table we can see  that  unlike  the OLS  methods,   the  EL  methods    give  closers   fits that are obtained from the   robust regression estimation  methods.  We   further   observe  from   Figure \ref{Fig1} that  among the EL    estimates the  robust EL estimates obtained from the Tukey   constraint   gives  the  best fit  to data avoiding  contaminated   points.   The  line obtained from   these  estimates   fits the  majority  of  the data points.

\section{Conclusion}

  The EL  method can be used instead of classical estimation like the OLS when there is no distribution assumption on error terms. The EL method maximizes a function  of  unknown probabilities corresponding to each observation under some constraints  related  to the unknown probabilities and the moments equations of the unknown parameters of interest. In the classical  EL  estimation for regression the constraints  related  to  the  parameters   of interest  are very similar to the normal equations in the OLS method. The only difference is: in OLS case we take average, however in EL case the constraints are a weighted form of normal equations obtained in the OLS method. Here, the weights are the unknown probabilities used in the EL function. However, it is well known that the OLS method  and hence the  corresponding  equations  have  poor performance when  there are some outliers in the data. Although, the observations in the constraints are weighted using the unknown probabilities,  these weights are not satisfactory to deal with the outliers. Therefore,   some  extra care should be  taken  to reduce the affect of outliers on the  estimation procedure.   In this paper, we  have  considered the  EL method with robustifyed constraints. The robustification has been  done by using some weight functions borrowed from robust M estimation. We  have  provided  a small simulation study and a real data example to demonstrate the capability of the robust EL method to handle unusual observations in the data. The simulation and the real data example results have showed  that   the  robust constraints have plausible affect on the  estimators  when heavy tailedness and/or outliers are possible in the data.


\newpage

\begin{table}
\centering
\caption{The MSE values of estimators }
\label{table1}
\begin{tabular}{llllll}
\hline

         k & n  & $ELM$ & $ELM_{TUK}$ & $ELM_{HUB}$ & $OLS$  \\ \hline
\multirow{3}{*}{2}
&30	& 0,1032& 0,1113& 0,1089& 0,0698\\
&50	& 0,0965& 0,1111& 0,1119& 0,0377\\
&100& 0,0670& 0,0688& 0,0718& 0,0207\\

                   \hline
\multirow{3}{*}{5}
&30	& 0,2126& 0,2107& 0,2090& 0,2083\\
&50	& 0,1323& 0,1266& 0,1283& 0,1008\\
&100& 0,1041& 0,0996& 0,1037& 0,0605\\

                   \hline
\multirow{3}{*}{15}
&30	&0,8651 &0,8502 &0,8643 &1,1643\\
&50	&0,4201 &0,4342 &0,4310 &0,4123\\
&100&0,2272 &0,2214 &0,2269 &0,1724\\

                   \hline
\end{tabular}\\
\end{table}

\begin{table}
\centering
\caption{The relative efficiencies of estimators with respect to the OLS }
\label{table2}
\begin{tabular}{lllll}
\hline
         k & n  & $ELM$ & $ELM_{TUK}$ & $ELM_{HUB}$  \\ \hline
\multirow{3}{*}{2}
&30	&1,4784 &1,5941 &1,5599 \\
&50	&2,5622 &2,9470	&2,9695	\\
&100&3,2329	&3,3193	&3,4628\\

                   \hline
\multirow{3}{*}{5}
&30	&1,0203 &1,0113 &1,0032\\
&50	&1,3117 &1,2557 &1,2726\\
&100&1,7208 &1,6469 &1,7137\\
                   \hline
\multirow{3}{*}{15}
&30	&0,7430 &0,7302 &0,7423\\
&50	&1,0190 &1,0532 &1,0455\\
&100&1,3175 &1,2839 &1,3165\\
                   \hline
\end{tabular}\\
\end{table}

\begin{table}
\centering
\caption{The MSE values of estimators with 10\% outlier}
\label{table3}
\begin{tabular}{llllll}
\hline

         k & n  & $ELM$ & $ELM_{HUB}$ & $ELM_{TUK}$ & $OLS$  \\ \hline
\multirow{3}{*}{2}
&30	&0.9490	&0.7183	&0.4338	&1.4127\\
&50	&1.1896	&0.6350	&0.2023	&1.3495\\
&100&0.9235	&0.6128	&0.2582	&1.1596\\

                   \hline
\multirow{3}{*}{5}
&30	&1.6607	&0.9600	&0.7361	&3.4963\\
&50	&1.3957	&1.0074	&0.4344	&2.0055\\
&100&1.1178	&0.9644	&0.4140	&1.4192\\

                   \hline
\multirow{3}{*}{15}
&30	&18.8479&12.7101&12.9962&48.4243\\
&50	&5.7769	&2.1413	&1.4580	&20.1266\\
&100&4.2138 &2.0062	&1.1010	&10.1531\\

                   \hline
\end{tabular}\\
\end{table}

\begin{table}
\centering
\caption{The relative efficiencies of estimators with respect to the OLS with 10\% outlier}
\label{table4}
\begin{tabular}{lllll}
\hline

         k & n  & $ELM$ & $ELM_{HUB}$ & $ELM_{TUK}$  \\ \hline
\multirow{3}{*}{2}
&30	&0.6718	&0.5085	&0.3071\\
&50	&0.8815	&0.4705	&0.0343\\
&100&0.7964	&0.5285	&0.2227\\
                   \hline
\multirow{3}{*}{5}
&30	&0.4750 &0.2746	&0.2105\\
&50	&0.6959 &0.5023	&0.2166\\
&100&0.7877 &0.6796	&0.2917\\
                   \hline
\multirow{3}{*}{15}
&30	&0.3892	&0.2625	&0.2678\\
&50	&0.2870 &0.1064	&0.0724\\
&100&0.4138	&2.1976	&0.1084\\
                   \hline
\end{tabular}\\
\end{table}

\newpage

\begin{table}
\centering
\caption{Number of International Calls from Belgium }\label{table5}
\fontsize{7.5pt}{7.2}\selectfont
\begin{tabular}{c c c c c c}
  \hline
   $Year$ & $Number of Calls^{a}$  & $Year$ & $Number of Calls^{a}$  & $Year$ & $Number of Calls^{a}$ \\
 $(x_{i})$ &$(y_{i})$&$(x_{i})$ &$(y_{i})$&$(x_{i})$ &$(y_{i})$\\
  \hline
50	&0,44 &58	&1,06 &66	&14,2\\
51	&0,47 &59	&1,2  &67	&15,9\\
52	&0,47 &60	&1,35 &68	&18,2\\
53	&0,59 &61	&1,49 &69	&21,2\\
54	&0,66 &62	&1,61 &70	&4,3\\
55	&0,73 &63	&2,12 &71	&2,4\\
56	&0,81 &64	&11,9 &72	&2,7\\
57	&0,88 &65	&12,4 &73	&2,9\\
\hline
\end{tabular}\\
\textit{$^{a}$In tens of millions.}
\end{table}

\begin{table}
\centering
\caption{Estimations for International Calls from Belgium }\label{table6}
\begin{tabular}{c c c c c}
  \hline
    & $EL$  & $EL_{TUK}$ & $EL_{HUB}$ & $OLS$\\
    \hline
 $\widehat{\beta}_{0}$& -6,7247 &-6,7254 &-10,8456 &-26,0059\\
 $\widehat{\beta}_{1}$&  0,1809 & 0,1378 &  0,2152 &0,5041\\
 \hline
\end{tabular}
\end{table}

\end{document}